\documentclass[12pt,preprint]{aastex}

\shorttitle{Revisiting X-ray Diagnostics}
\shortauthors{Porter \& Ferland}

\begin{document}
\title{Revisiting He-like X-ray Emission Line Plasma Diagnostics}


\author{R. L. Porter \& G. J. Ferland}
\affil{Dept. of Physics and Astronomy, University of Kentucky, Lexington, KY, 40506}
\email{rporter@pa.uky.edu}

\begin{abstract}
A complete model of helium-like line and continuum emission has been incorporated into the plasma simulation code Cloudy. All elements between He and Zn are treated, any number of levels can be considered, and radiative and collisional processes are included.  This includes photoionization from all levels, line transfer including continuum pumping and destruction by background opacities, scattering, and collisional processes.  The model is calculated self-consistently along with the ionization and thermal structure of the surrounding nebula.  The result is a complete line and continuum spectrum of the plasma.  Here we focus on the ions of the He~I sequence and reconsider the standard helium-like X-ray diagnostics.  We first consider semi-analytical predictions and compare these with previous work in the low-density, optically-thin limit.  We then perform numerical calculations of helium-like X-ray emission (such as is observed in some regions of Seyferts) and predict line ratios as a function of ionizing flux, hydrogen density, and column density.  
In particular, we demonstrate that, in photoionized plasmas, the $R$-ratio, a density indicator in a collisional plasma, depends on the ionization fraction and is strongly affected by optical depth for large column densities.  We also introduce the notion that the $R$-ratio is a measure of the incident continuum at UV wavelengths.  The $G$-ratio, which is temperature-sensitive in a collisional plasma, is also discussed, and shown to be strongly affected by continuum pumping and optical depth as well.  These distinguish a photoionized plasma from the more commonly studied collisional case.
\end{abstract}


\section{Introduction}
X-ray emission lines from high-excitation ions offer a different view of Active Galactic Nuclei (AGN) emission line regions than do the strong UV and optical lines that are usually studied. While emission lines such as H~I Ly$\alpha$, He~I $\lambda5876$, and C~IV $\lambda$1549 come from gas with electron temperature $T_e \sim 10^4$K, the emission lines which are detected in X-ray spectra can come from gas that is much hotter and more highly ionized, and which presumably lies closer in to the ionizing continuum source. Such X-ray emission lines are often discussed in connection with observations of X-ray and UV warm absorbers (eg. Kraemer et al. 2005; Netzer et al.2003; George et al. 1995). Emission lines from helium-like ions are important in this regard, because the ratios of their X-ray lines provide temperature and density diagnostics that can be measured from intermediate-resolution X-ray spectra. Several of these lines have now been detected in a number of nearby AGN, including Mrk 279 (Costantini et al. 2005; Kaastra et al. 2004), NGC 5548 (Steenbrugge et al. 2005), NGC 3783 (Netzer et al. 2003; Kaspi et al. 2002), NGC 4051 (Collinge et al. 2001), NGC 7469 (Blustin et al. 2003), and NGC 4151 (Kraemer et al. 2005).

The X-ray forbidden (f) and intercombination (i) lines connect the triplet states to the singlet ground state, while the resonance (r) line is due to transitions into the ground state from $2p\,{}^{1}\!P$.  (See the schematic energy-level diagram in Figure~\ref{fig:grotrian2}).
Diagnostic line ratios for obtaining temperature, $G(T_{\mathrm{e}})=(I_{\mathrm{f}}+I_{\mathrm{i}})/I_{\mathrm{r}}$, and density, $R(n_{\mathrm{e}})=I_{\mathrm{f}}/I_{\mathrm{i}}$,  from particular X-ray line ratios in plasmas have been discussed by a number of authors (Bautista \& Kallman 2000; Porquet \& Dubau 2000; Porquet et al. 2001; Pradhan \& Shull 1981; Pradhan 1985; Blumenthal, Drake, \& Tucker 1972; and many others).  These were introduced by Gabriel \& Jordan (1969, 1973) and were originally developed for coronal plasmas, not where photoionization-recombination is dominant. Recently, authors have discussed the influence of radiative transfer effects on the standard X-ray diagnostics (see Coup\'e et al. 2004, Godet et al. 2004, and Bianchi \& Matt 2002).  

The theoretical models of the He~I isoelectronic sequence presented here have been developed as part of the plasma simulation code Cloudy (last described by Ferland et al. 1998).  The model helium atom has been discussed by Porter et al. (2005, 2007) and Bauman et al. (2005).  Where the sources and methods used in the model ions differ from those previously discussed sources and methods of the model helium atom, those differences are given in Appendix~~\ref{AppendixData}.  

Here we consider the standard X-ray diagnostics from both semi-analytical and numerical perspectives.  In section~\ref{semianalytical}, we derive semi-analytical expressions of line ratios in the limit where the lines are formed by recombination and calculate the line ratios for helium-like oxygen and iron at appropriate temperatures.  We compare our results with those of other workers and discuss how radiative transfer and collisions will affect the emission.   In section~\ref{numerical}, we include all relevant physics in a numerical model to calculate theoretical line ratios as functions of column density.  This illustrates the effect of increasing optical depth.  We also calculate helium-like emission as a function of ionizing flux and hydrogen density.  We present the predicted line ratios for oxygen and discuss how our results affect the interpretaion of existing and future measurements of the emission lines from AGN.  We also discuss the effects of continuum pumping.  We conclude in section~\ref{conclusions}.

\section{Semi-analytical Calculations}
\label{semianalytical}
Gabriel \& Jordan (1969, 1973) showed how ratios of intensities of various lines involving transitions between the $n=2$ and $n=1$ levels of helium-like ions could be used to measure temperature and density in a collisionally-ionized plasma.  In such a gas the electron kinetic energy is roughly equal to the ionization potentials of the species that are present.   Collisional excitation of the $n=2$ levels from ground is the main process that produces the X-ray lines.  

These results do not carry over to the photoionized plasmas considered here, however.  In this second case the ionization and kinetic temperature are controlled by the radiation field.  The kinetic energy is much less than the ionization potentials of the dominant stages of ionization and the $n=2$ levels are mainly populated by recombination from the next higher ionization stage.  As a result the $R$ and $G$ ratios are not density and temperature indicators, but, as we show here, are determined by the ionization and column density.  

In the following we will consider helium-like oxygen (recombination onto O$^{7+}$ forming O$^{6+}$) and iron (recombination onto Fe$^{25+}$ forming Fe$^{24+}$) in detail.  Tests show that, for an AGN continuum and solar abundances, the kinetic temperature is roughly 500,000~K and $10^{7}$~K for these two ions.  We will assume these temperatures in the estimates that follow.

\subsection{Level Populations}
\label{levpops}

\begin{figure}
\centering
\includegraphics[width=5.0in,keepaspectratio=true]{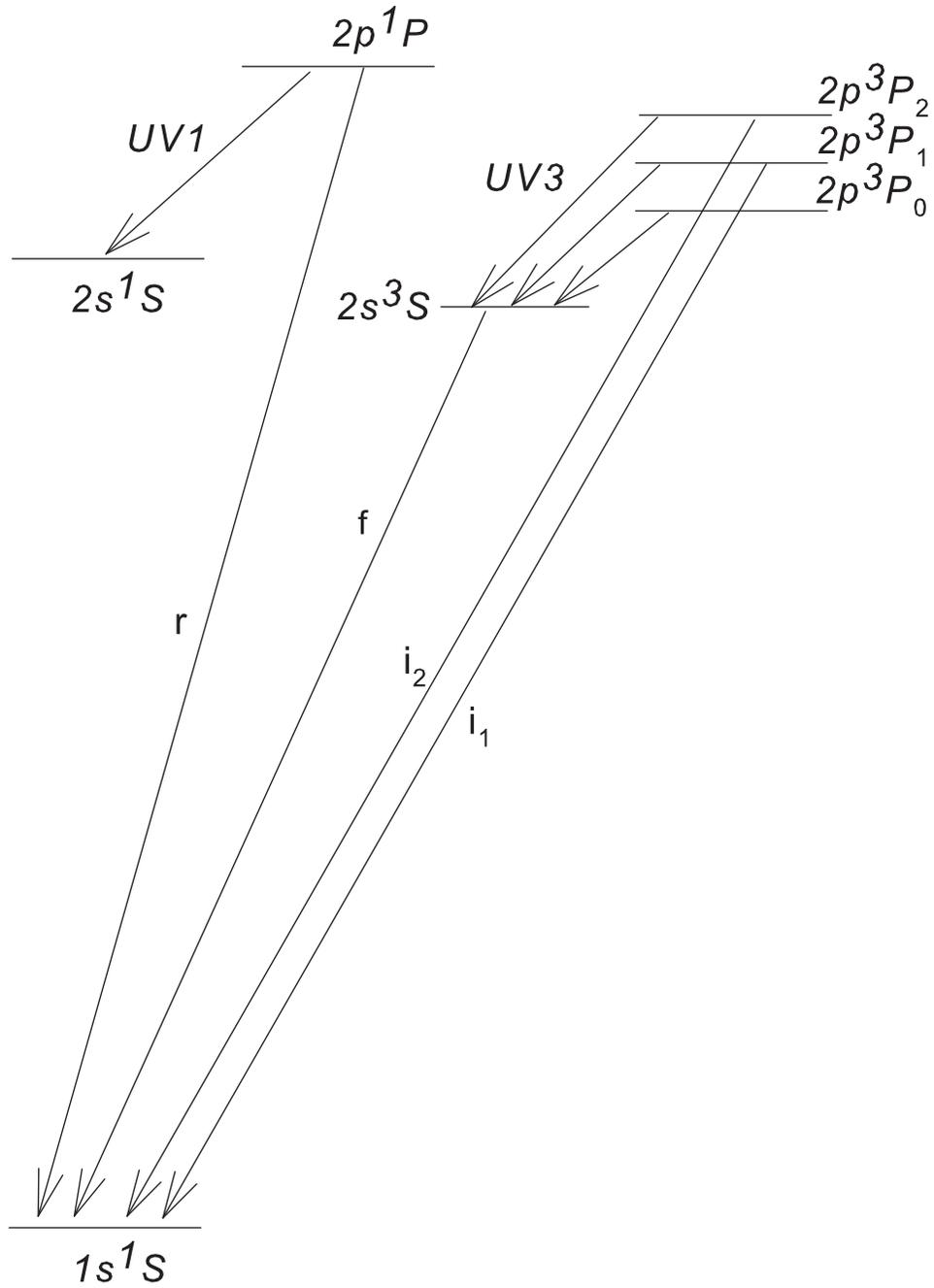}
\caption[Grotrian Diagram of Helium-like Oxygen.]{Grotrian diagram of the $n=1$ and $n=2$ levels of helium-like oxygen.  Some transitions are not shown.  Relative energies are not drawn to scale.  The energy order is different for different ions of the sequence.}
\label{fig:grotrian2}
\end{figure}

The theoretical intensities of our lines of interest (transitions from $n=2$ to the ground state) depend on the populations of the upper levels of each transition: $2p\,{}^{3}\!P$, $2s\,{}^{3}\!S$, and $2p\,{}^{1}\!P$.  In a low-density, pure recombination plasma, the time-steady balance equations for the populations of individual levels involve only effective recombination coefficients and transition probabilities.  We consider that limit here.  We define $\alpha^{eff}_i(Z,T)$ as the effective recombination coefficient (including dielectronic recombination) into level $i$ of the He-like ion of nuclear charge $Z$ at temperature $T$.  We calculate $\alpha^{eff}_i(Z,T)$ with the use of cascade probabilities, as outlined by Robbins (1968).  (See also Osterbrock \& Ferland 2006, page 85 for a discussion of effective recombination.)  We include $nLS$-resolved terms with principal quantum number $n\leq20$ and a series of `collapsed' $n$-resolved levels with $20\leq n\leq100$ (as discussed in Porter et al. 2005 and Bauman et al. 2005).  See Appendix~\ref{AppendixData} for a discussion of our atomic sources, and Figure~\ref{fig:grotrian2} for an energy level diagram.  Note that in Figure~\ref{fig:grotrian2} and in the text, we refer to the UV lines $2p\,{}^{3}\!P_j-2s\,{}^{3}\!S$ and $2p\,{}^{1}\!P-2s\,{}^{1}\!S$ as UV3$_j$ and UV1, respectively.
The individual components of the intercombination doublet $2p\,{}^{3}\!P_j-1s\,{}^{1}\!S$ are designated by i$_j$. 

Ions can radiatively decay from the $2p\,{}^{3}\!P$ term via the electric-dipole transition $2p\,{}^{3}\!P-2s\,{}^{3}\!S$ and via the intercombination doublet $2p\,{}^{3}\!P_j-1s\,{}^{1}\!S$. 
The intercombination transition probability, $A_{2p\,{}^{3}\!P_1-1s\,{}^{1}\!S}(Z)$, (approximately $\propto Z^8$)
increases with increasing nuclear charge, $Z$, faster
than the electric-dipole transition probability, $A_{2p\,{}^{3}\!P-2s\,{}^{3}\!S}(Z)$, (approximately $\propto Z^2$) so that, for $Z>6$, the intercombination line is stronger than the electric-dipole transition.  The transition probabilities to ground from the separate $2p\,{}^{3}\!P_j$ levels are vastly different.  
The transition $2p\,{}^{3}\!P_0-1s\,{}^{1}\!S$ is strictly forbidden (for one-photon transitions by the selection rule forbidding $J=0\Leftrightarrow0$, and for two-photon transitions by the rule requiring $\Delta L=0,2$)
while the transition probability $A_{2p\,{}^{3}\!P_1-1s\,{}^{1}\!S}(Z)$ is the fastest, about a thousand times faster than $A_{2p\,{}^{3}\!P_2-1s\,{}^{1}\!S}(Z)$ for carbon and nearly 7000 times faster for iron.  These large differences in transition probabilities make it necessary to solve for the populations of the separate $j$-levels separately.  The transition probabilities $A_{2p\,{}^{3}\!P_j-2s\,{}^{3}\!S}(Z)$ also depend upon $j$.  These differences are only as large as about $25\%$ for carbon, while in the case of iron the transition probability for $j=2$ is 3-4 times the $j=0$ and $1$ values.  The populations, $n_{2p\,{}^{3}\!P_j}$, of the 
$2p\,{}^{3}\!P_j$ terms\footnote[1]{The $ST$-mixing discussed in Bauman et al. 2005 for helium becomes important for the high $Z$ ions of the sequence as the physical $2p\,{}^{3}\!P_1$ level is mixed roughly 30\% with the $LS$ $2p\,{}^{1}\!P$ level in the case of iron.  We do not consider this effect here.  See Lin et al. (1977).} in the low density limit are as follows:
\begin{eqnarray}
\label{eqn:2p3P0pops}
\frac{n_{2p\,{}^{3}\!P_0}(Z,T)}{n_e~n_{\mathrm{Z}^+}} &=& \frac{g_0~\alpha^{eff}_{2p\,{}^{3}\!P_0}(Z,T)}{A_{2p\,{}^{3}\!P_0-2s\,{}^{3}\!S}(Z)}\ \ [\mathrm{cm}^{3}]\\
\label{eqn:2p3P1pops}
\frac{n_{2p\,{}^{3}\!P_1}(Z,T)}{n_e~n_{\mathrm{Z}^+}} &=& \frac{g_1~\alpha^{eff}_{2p\,{}^{3}\!P_0}(Z,T)}{\left[A_{2p\,{}^{3}\!P_1-2s\,{}^{3}\!S}(Z)+A_{2p\,{}^{3}\!P_1-1s\,{}^{1}\!S}(Z)\right]}\ \  [\mathrm{cm}^{3}]\\
\label{eqn:2p3P2pops}
\frac{n_{2p\,{}^{3}\!P_2}(Z,T)}{n_e~n_{\mathrm{Z}^+}} &=& \frac{g_2~\alpha^{eff}_{2p\,{}^{3}\!P_0}(Z,T)}{\left[A_{2p\,{}^{3}\!P_2-2s\,{}^{3}\!S}(Z)+A_{2p\,{}^{3}\!P_2-1s\,{}^{1}\!S}(Z)\right]}\ \ [\mathrm{cm}^{3}],
\end{eqnarray}
where $n_{Z^+}$ is the density of the hydrogen-like ion, $T$ is the electron temperature, $n_e$ is the electron density, and we have used $\alpha^{eff}_{2p\,{}^{3}\!P_j}(Z,T)=g_j~\alpha^{eff}_{2p\,{}^{3}\!P_0}(Z,T)$ and $A_{2p\,{}^{3}\!P_0-1s\,{}^{1}\!S}(Z)=0$.  

Ions can radiatively decay from the $2p\,{}^{1}\!P$ level to either the ground state, $2s\,{}^{1}\!S$, or $2s\,{}^{3}\!S$.  The decay to ground dominates for all $Z$ but is highly susceptible to optical depth effects since it is a resonance line - this is discussed more below.  The rate of the transition to $2s\,{}^{3}\!S$ (approximately $\propto Z^7$) increases with increasing $Z$ faster
than the decay to $2s\,{}^{1}\!S$ (approximately $\propto Z^2$) so that the two transition probabilities are comparable in the case of helium-like iron.  The population, $n_{2p\,{}^{1}\!P}$, of the $2p\,{}^{1}\!P$ term is
\begin{equation}
\frac{n_{2p\,{}^{1}\!P}(Z,T)}{n_e~n_{\mathrm{Z}^+}} = \frac{\alpha^{eff}_{2p\,{}^{1}\!P}(Z,T)}{\left[A_{2p\,{}^{1}\!P-1s\,{}^{1}\!S}(Z)+A_{2p\,{}^{1}\!P-2s\,{}^{1}\!S}(Z)+A_{2p\,{}^{1}\!P-2s\,{}^{3}\!S}(Z)\right]}\ \ [\mathrm{cm}^{3}].
\label{eqn:2p1Ppop}
\end{equation}

Ions can radiatively decay from the metastable $2s\,{}^{3}\!S$ level to the ground state via two-photon and magnetic-dipole transitions.  The magnetic-dipole decay dominates for all $Z$, and we accordingly ignore the two-photon decay in the semi-analytical discussion.  The population, $n_{2s\,{}^{3}\!S}$, of the $2s\,{}^{3}\!S$ level is given by 
\begin{equation}
\frac{n_{2s\,{}^{3}\!S}(Z,T)}{n_e~n_{\mathrm{Z}^+}} = \frac{\alpha^{eff}_{2s\,{}^{3}\!S}(Z,T)}{A_{2s\,{}^{3}\!S-1s\,{}^{1}\!S}(Z)}\ \ [\mathrm{cm}^{3}].
\label{eqn:2s3Spop}
\end{equation}

Equations \ref{eqn:2p3P0pops}-\ref{eqn:2s3Spop} are valid when triplet-singlet exchange collisions (which change the spin of one of the electrons) can be neglected.
To first order, exchange collisions add an additional term in each equation.  Equation~\ref{eqn:2s3Spop}, for example, would be modified as follows: 
\begin{equation}
\frac{n_{2s\,{}^{3}\!S}(Z,T)}{n_e~n_{\mathrm{Z}^+}} = \frac{\alpha^{eff}_{2s\,{}^{3}\!S}(Z,T)}{A_{2s\,{}^{3}\!S-1s\,{}^{1}\!S}(Z)+\sum\limits_{n,L}n_e~q_{2s\,{}^{3}\!S-nl\,{}^{1}\!L}(Z,T)}\ \ [\mathrm{cm}^{3}].
\label{eqn:exchange}
\end{equation}
where $q_{2s\,{}^{3}\!S-nl\,{}^{1}\!L}(Z,T)$ is the collision rate coefficient (in units cm$^{3}$~s$^{-1}$)
from $2s\,{}^{3}\!S$ to the singlet term $nl\,{}^{1}\!L$.  We define $q_{tot}(Z,T) = \sum\limits_{n,L}q_{2s\,{}^{3}\!S-nl\,{}^{1}\!L}(Z,T)$.  Collisions are negligible if $n_e~q_{tot}(Z,T)$ is small relative to $A_{2s\,{}^{3}\!S-1s\,{}^{1}\!S}(Z)$.  Considering oxygen at 500,000~K, a temperature near where the hydrogen-like ionization state peaks, we find $A_{2s\,{}^{3}\!S-1s\,{}^{1}\!S}(6)=10^{3}$~s$^{-1}$ and $q_{tot}(8,10^{5.7}$~K$)\approx4\times10^{-10}$~cm$^{3}$~s$^{-1}$, corresponding to a critical electron density of $2\times10^{12}$~cm$^3$.  For iron at a temperature of $10^{7}$~K, we find a critical density of $2\times10^{19}$~cm$^3$.

Because the transition probabilities (approximately $\propto Z^{10}$) increase with increasing $Z$, while the collisional rate coefficients decrease with increasing transition energies (approximately $\propto Z^2$), the critical density will be even larger for heavier ions of the sequence.  We restrict our semi-analytical calculations to lesser densities and neglect exchange collisions.  

\subsection{Emissivities}
\label{emissivities}

The total emission, $4\pi~j_{\lambda}/n_e~n_{\mathrm{Z}^+}$, of a recombination line with wavelength $\lambda$ is 
\begin{equation}
\frac{4\pi~j_{\lambda}}{n_e~n_{\mathrm{Z}^+}} = \frac{hc}{\lambda}~\frac{n_u(Z,T)}{n_e~n_{\mathrm{Z}^+}}~A_{ul}(Z)\ \ [\mathrm{ergs~cm}^{3}~\mathrm{s}^{-1}]
\label{eqn:emiss}
\end{equation}
where $A_{ul}(Z)$ is the transition probability of the transition and Z$^+$ is the parent (hydrogenic) ion.  The dimensionless ratios of the forbidden and resonance emissivities to the intercombination emissivity are as follows:
\begin{eqnarray}
\label{eqn:Eratiof}
R=\frac{j_{\mathrm{f}}(Z,T)}{j_{\mathrm{i}}(Z,T)} &=& \frac{\lambda_{\mathrm{i}}}{\lambda_{\mathrm{f}}}~\left[\frac{A_{2s\,{}^{3}\!S-1s\,{}^{1}\!S}(Z)}{A_{2p\,{}^{3}\!P_1-1s\,{}^{1}\!S}(Z)}~\frac{n_{2s\,{}^{3}\!S}(Z,T)}{n_{2p\,{}^{3}\!P_1}(Z,T)}+~\frac{A_{2s\,{}^{3}\!S-1s\,{}^{1}\!S}(Z)}{A_{2p\,{}^{3}\!P_2-1s\,{}^{1}\!S}(Z)}~\frac{n_{2s\,{}^{3}\!S}(Z,T)}{n_{2p\,{}^{3}\!P_2}(Z,T)}\right]\\
\label{eqn:Eratior}
L=\frac{j_{\mathrm{r}}(Z,T)}{j_{\mathrm{i}}(Z,T)} &=& \frac{\lambda_{\mathrm{i}}}{\lambda_{\mathrm{r}}}~\left[\frac{A_{2p\,{}^{1}\!P-1s\,{}^{1}\!S}(Z)}{A_{2p\,{}^{3}\!P_1-1s\,{}^{1}\!S}(Z)}~\frac{n_{2p\,{}^{1}\!P}(Z,T)}{n_{2p\,{}^{3}\!P_1}(Z,T)}+\frac{A_{2p\,{}^{1}\!P-1s\,{}^{1}\!S}(Z)}{A_{2p\,{}^{3}\!P_2-1s\,{}^{1}\!S}(Z)}~\frac{n_{2p\,{}^{1}\!P}(Z,T)}{n_{2p\,{}^{3}\!P_2}(Z,T)}\right]
\end{eqnarray}
where $\lambda_{\mathrm{i}}$, $\lambda_{\mathrm{f}}$, and $\lambda_{\mathrm{r}}$ are the wavelengths of the intercombination, forbidden, and resonance transitions, respectively and we have given the label ``$L$'' to the ratio of resonance to intercombination emission.
The $G$-ratio, defined as the sum of the intercombination and forbidden emission divided by the resonance emission, is given by
\begin{equation}
G=\frac{j_{\mathrm{f}}(Z,T)+j_{\mathrm{i}}(Z,T)}{j_{\mathrm{r}}(Z,T)}=\frac{R+1}{L} \\
\label{eqn:EratioG}
\end{equation}


\subsection{Comparison with Previous Work}
\label{oxyexample}

The results we obtain from the above analyses are in good agreement with previous work. At a temperature of 500,000~K, typical of a plasma in which recombination onto helium-like oxygen is important, we find from Equation~\ref{eqn:Eratiof} the O~VII ratio $R=j_{\mathrm{f}}/j_{\mathrm{i}}=4.2$.  At the same temperature, Figure~8 of Porquet \& Dubau (2000) indicates $R\approx4.3$ in the low-density limit, while Figure~3 of Bautista \& Kallman (2000) suggests $R\approx4.2$.  The value calculated here agrees very well with those two values.  We find from Equation~\ref{eqn:EratioG} the ratio  $G=(j_{\mathrm{f}}+j_{\mathrm{i}})/j_{\mathrm{r}}=5.0$.  Porquet \& Dubau find (as taken from their Figure~7) $G\approx4.8$ when the hydrogen-like ionization stage of oxygen is at its peak.  Figure~4 of Bautista \& Kallman reports $G\approx5.0$ at the conditions considered here.  Again, the present value agrees very well with values found by other workers.  

For the case of iron, at a temperature of $10^{7}$~K, we find $G=4.2$ and $R=0.68$.  The results from the Bautista \& Kallman (2000) work are $G=4.7$ and $R=0.71$ (Bautista, private communication).  These results are also in good agreement with our results.

\subsection{Optical Depth Effects}
\label{opticaldepth}

Introducing escape probabilities, which modify transition probabilities to yield \textit{effective} transition probabilities, the emission per unit volume is given by 
\begin{equation}
4\pi~I_{\lambda} = \frac{hc}{\lambda}~n_u(Z,T)~A_{ul}(Z)~\epsilon_{\lambda}\ \ [\mathrm{ergs~cm}^{-3}~\mathrm{s}^{-1}].
\label{eqn:intens}
\end{equation}
(See Elitzur 1992, for a discussion of escape probabilities.)  Small optical depth corresponds to an $\epsilon$ of unity, and $\epsilon\approx\tau^{-1}$ when $\tau$ is large.  The optical depth of a line increases as the column density of the ion increases but also depends upon the wavelength and transition probability of the line.  The effect of optical depth on line ratios is dependent upon which line becomes optically thick first (or which escape probability becomes less than unity first) as column density increases.  It is important to note, however, that, in addition to the explicit escape probability dependence in Equation~\ref{eqn:intens}, there is also an implicit dependence in the population $n_u(Z,T)$\footnote[2]{Actually, effective recombinations can also be affected by line optical depths.  This is included in the numerical calculations below but is a complication beyond the scope of these semi-analytical calculations.}.  For an upper level with only one significant decay mode, these escape probabilities cancel in the low density limit (see, for example, $n_{2s\,{}^{3}\!S}(Z,T)$ given in Equation~\ref{eqn:2s3Spop}), and the line is said to be effectively optically thin.  For an upper level with more than one significant decay mode, the escape probabilities do not cancel out, so line emission from these levels is affected by optical depth.  For example, the ratio of the forbidden to i$_1$ intensities (equal to the $R$-ratio where the i$_2$ line can be neglected, as with O~VII) varies as follows
\begin{equation}
\frac{f}{i_1} \propto \frac{A_{2p\,{}^{3}\!P_1-1s\,{}^{1}\!S}(Z)~\epsilon_{\mathrm{i_1}}+A_{2p\,{}^{3}\!P_1-2s\,{}^{3}\!S}(Z)~\epsilon_{\mathrm{UV3}_1}}{A_{2p\,{}^{3}\!P_1-1s\,{}^{1}\!S}(Z)~\epsilon_{\mathrm{i_1}}},
\label{eqn:r_propto}
\end{equation}
where $\epsilon_{\mathrm{i}_1}$ and $\epsilon_{\mathrm{UV3}_1}$ are the escape probabilities of the i$_1$ and UV3$_1$ lines, respectively.  See Appendix~\ref{AppendixAbsorption} for a demonstration that the intercombination line can become optically thick.  

The intercombination rate dominates in Equation~\ref{eqn:r_propto} for all $Z>6$.  The intercombination line also becomes optically thick well before the UV3$_1$ line.  (Optical depth in the UV3$_j$ lines is also discussed in Appendix~\ref{AppendixAbsorption}.) For oxygen at 500,000~K, the $R$-ratio begins to be affected by optical depth at about $N(\mathrm{O}^{6+})=10^{22}$~cm$^{-2}$.  When this happens, as we will see below, $R$ will surpass the classical, canonical value as the denominator of the ratio is surpressed.  For iron, the intercombination lines will not be affected by optical depth at all (this is also discussed below).  The fact that the intercombination line can become optically thick was mentioned by Godet et al. (2004) but was not demonstrated or indicated in any results.     

Recent observational results on the $G$ and $R$-ratios in NGC~4151 can be found in Armentrout et al. (2007).  Kinkhabwala et al. (2002) presented large observed values of the $R$-ratio (in NGC 1068).  The present work demonstrates that optical thickness in the intercombination line is a possible explanation.

While the $G$-ratio can depend strongly on optical depth, it is also strongly affected by direct continuum pumping of the resonance line.  The transition with increasing column density from the pumped case to the optically thick case is not well treated with an escape probability formalism.  Accordingly, these effects on the $G$-ratio will be neglected in the semi-analytical analysis, although they are included in the numerical solutions.

\subsection{Individual Components of the Intercombination Doublet}
\label{imultiplet}
Here we present a method by which the optical depth of intercombination lines, and so the column density of its ion, can be derived from the relative intensities of the individual components of the intercombination doublet.  

The individual components i$_j$ of the intercombination doublet are related by the following expression:
\begin{equation}
\frac{I_{\mathrm{i_2}}(Z,T)}{I_{\mathrm{i_1}}(Z,T)} = \frac{\lambda_{\mathrm{i_1}}}{\lambda_{\mathrm{i_2}}}~\frac{g_2}{g_1}~\frac{A_{2p\,{}^{3}\!P_2-1s\,{}^{1}\!S}(Z)}{A_{2p\,{}^{3}\!P_1-1s\,{}^{1}\!S}(Z)\epsilon_{\mathrm{i_1}}}~\left[\frac{A_{2p\,{}^{3}\!P_1-2s\,{}^{3}\!S}(Z)\epsilon_{\mathrm{UV3_1}}+A_{2p\,{}^{3}\!P_1-1s\,{}^{1}\!S}(Z)\epsilon_{\mathrm{i_1}}}{A_{2p\,{}^{3}\!P_2-2s\,{}^{3}\!S}(Z)\epsilon_{\mathrm{UV3_2}}+A_{2p\,{}^{3}\!P_2-1s\,{}^{1}\!S}(Z)}\right],
\label{eqn:i1i2} 
\end{equation}
where $\epsilon_{\mathrm{UV3_j}}$ and $\epsilon_{\mathrm{UV1}}$ are the escape probabilities for the UV transitions $2p\,{}^{3}\!P_j-2s\,{}^{3}\!S$ and $2p\,{}^{1}\!P-2s\,{}^{1}\!S$, respectively, and $\epsilon_{\mathrm{i_1}}$ is the escape probability for the i$_1$ line.  We have not included escape probabilities for the i$_2$ line because that line will always have an optical depth roughly three orders of magnitude less than the optical depth of the $i_1$ line.  The rightmost factor in Equation~\ref{eqn:i1i2} cannot be simplified for a general $Z$, because in both the numerator and the denominator the dominant transition probability is a function of $Z$.  Again, for the case of oxygen, we have
\begin{eqnarray}
\frac{I_{\mathrm{i_2}}(8,T)}{I_{\mathrm{i_1}}(8,T)} &\approx&   \frac{21.81}{21.8}~\frac{5}{3}~\frac{3.5\times10^5}{5.5\times10^8\epsilon_{\mathrm{i_1}}}~\left[\frac{8.1\times10^7\epsilon_{\mathrm{UV3_1}}+5.5\times10^8\epsilon_{\mathrm{i_1}}}{8.4\times10^7\epsilon_{\mathrm{UV3_2}}+3.5\times10^5}\right] \nonumber \\
\label{eqn:i1i2oxy} 
&\approx& \left(0.0011/\epsilon_{\mathrm{i_1}}\right)~\left[\frac{8.1\times10^7\epsilon_{\mathrm{UV3_1}}+5.5\times10^8\epsilon_{\mathrm{i_1}}}{8.4\times10^7\epsilon_{\mathrm{UV3_2}}+3.5\times10^5}\right].
\end{eqnarray}
If optical depth effects are not important, $I_{\mathrm{i_2}}/I_{\mathrm{i_1}}\approx0.0080$.  If optical depths are significant (but still small enough that the $3.5\times10^5$ in the denominator in the rightmost factor can be neglected), we have 
\begin{equation}
\frac{I_{\mathrm{i_2}}(8,T)}{I_{\mathrm{i_1}}(8,T)} \approx \frac{0.0011\epsilon_{\mathrm{UV3_1}}}{\epsilon_{\mathrm{UV3_2}}\epsilon_{\mathrm{i_1}}}+\frac{0.0069}{\epsilon_{\mathrm{UV3_1}}},
\end{equation}
and the ratio will increase with increasing optical depth.  If optical depth in the UV3$_j$ lines is not important (see Appendix~\ref{AppendixAbsorption}), the escape probability (inversely proportional to the column density) in the intercombination line can be derived from the individual components of the intercombination doublet as follows:
\begin{equation}
\epsilon_{\mathrm{i_1}} = \frac{0.0011}{I_{\mathrm{i_2}}/I_{\mathrm{i_1}}-0.0069} = f(N).
\label{eqn:epsiloni1oxy}
\end{equation}
A similar but different expression holds for other ions of the sequence.  For iron, the i$_1$ rate is so dominant that it cancels out completely in Equation~\ref{eqn:i1i2}, and the ratio is
\begin{equation}
\frac{I_{\mathrm{i_2}}(26,T)}{I_{\mathrm{i_1}}(26,T)} \approx \frac{5}{3} \left[\frac{1}{0.24~\epsilon_{\mathrm{UV3_2}} + 1}\right].   
\label{eqn:i1i2fe}
\end{equation}
The ratio therefore ranges only slightly, varying from about 4/3 for optically thin media to 5/3 for optically thick media.  With sufficiently high signal-to-noise observations, the optical depth of the UV3$_2$ line could be derived from X-ray observations of the intercombination lines. 

It is important to note that this analysis is largely independent of density because collisions will only begin to affect the relative populations of the separate $2p\,{}^{3}\!P_j$ levels for densities greater than $n_{e}=10^{14}$~cm$^{-3}$ for oxygen (at $500,000$~K) and $n_{e}=10^{16}$~cm$^{-3}$ for iron (at $10^{7}$~K). 

Resolving the individual components of the intercombination doublet may be difficult.  The line separation of the doublet corresponds to a velocity range from only about $20$~km~s$^{-1}$ for carbon to almost $700$~km~s$^{-1}$ for iron.  If the intrinsic line broadening is small, then the intercombination line can be detected in systems with significantly less broadening.  The individual components could be resolved with a spectrometer having resolving power of at least $12000$ for carbon and $400$ for iron.  Current and planned X-ray satellite observatories might be able to resolve the individual components for ions as light as magnesium and almost certainly for ions as heavy as iron.  Unfortunately, optical depth in the intercombination line is less likely for heavy ions because they are generally less abundant, and in practice the method outlined here may be untenable.  The best candidates for resolving the individual components of an optically thick intercombination doublet are probably Mg~XI, Si~XIII, and S~XV.


\section{Numerical Calculations}
\label{numerical}
\subsection{Summary}
Having established our semi-analytical results in the low-density, optically-thin limit, we now consider numerical calculations.  In this section, we present the results of Cloudy models of helium-like X-ray emission for a wide range of physical conditions.  These models span the range of emitters that may be found in high-ionization regions of AGN.  The emergent spectrum is calculated self-consistently with the ionization and thermal structure of the line-emitting region.  This includes collisional processes, line transfer, background opacities, continuum pumping, and dielectronic recombination.  

We first consider line ratios for a particular model as a function of hydrogen column density.  Then we vary the ionizing flux and volume density of the original model and present line ratios as contour plots.  Finally, we illustrate the effects of continuum pumping by adding an additional incident continuum source.

\subsection{Line Ratios as a Function of Column Density}
We calculate a constant density plane-parallel slab with the Korista et al. (1997) ionizing continuum incident with hydrogen-ionizing flux $\Phi(\mathrm{H})=10^{18}$~photons~cm$^{-2}$~s$^{-1}$ and hydrogen density $n_{\mathrm{H}}=10^7$~cm$^{-3}$, corresponding to $\log~U(\mathrm{H})\approx0.5$.  These parameters were chosen because helium-like oxygen emission peaks near these conditions for a small column density.  Solar abundances are assumed (Grevesse \& Sauval 1998; Allende Prieto et al. 2001, 2002; Holweger 2001).

\begin{figure}
\centering
\includegraphics[width=6.0in,keepaspectratio=true]{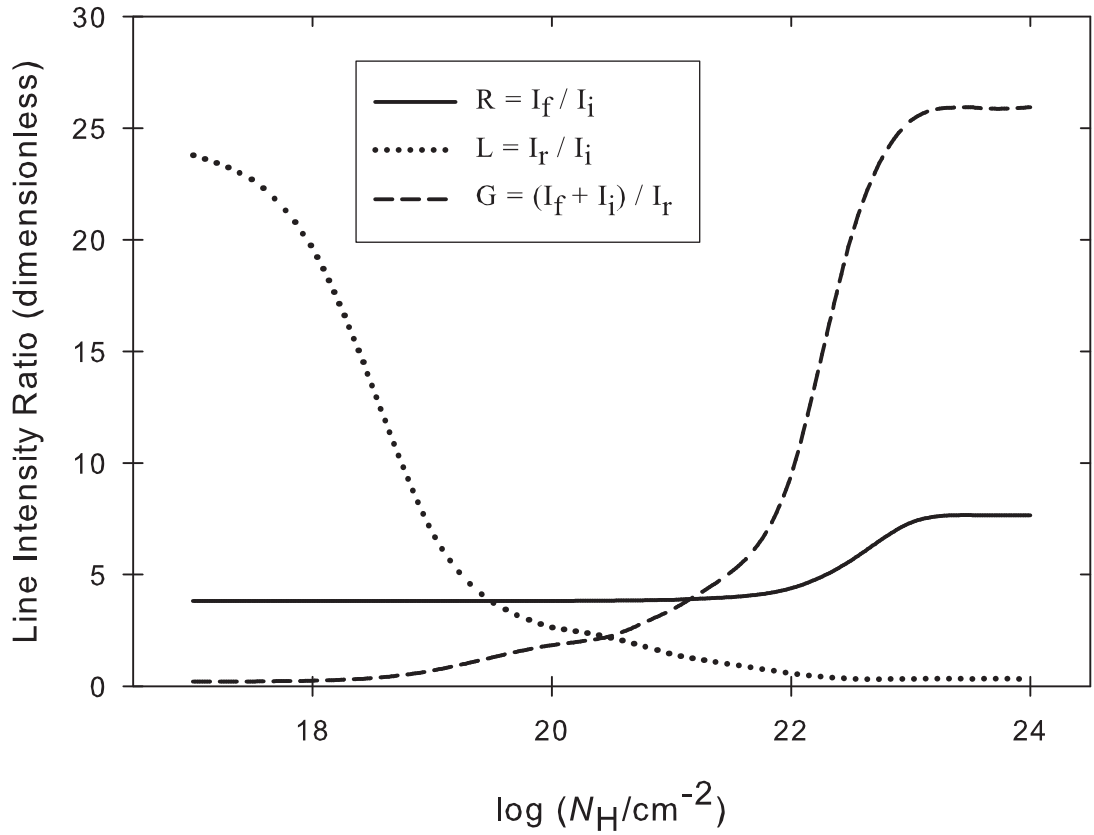}
\caption[$R=I_{\mathrm{f}}/I_{\mathrm{i}}$, $L=I_{\mathrm{r}}/I_{\mathrm{i}}$, and $G=(I_{\mathrm{f}}+I_{\mathrm{i}})/I_{\mathrm{r}}$ as a Function of Hydrogen Column Density.]{$R=I_{\mathrm{f}}/I_{\mathrm{i}}$, $L=I_{\mathrm{r}}/I_{\mathrm{i}}$, and $G=(I_{\mathrm{f}}+I_{\mathrm{i}})/I_{\mathrm{r}}$ as a function of hydrogen column density.  See text for details of model.}
\label{fig:randg}
\end{figure}

In Figure~\ref{fig:randg}, we plot the O~VII $R$, $L$, and $G$ ratios as a function of column density.  The $R$-ratio increases by a factor of $2$ at large column densities.  This is due to an increase in the optical depth of the intercombination line (see Equation~\ref{eqn:r_propto}) and suggests that caution should be used in deriving electron density from the $R$-ratio in conditions where the column density is large and the intercombination line may be optically thick.  For low column densities, $R$ agrees well with the semi-analytical value calculated above.  The $L$-ratio ($I_{\mathrm{r}}/I_{\mathrm{i}}$), however, falls by a factor of nearly $70$ over the same range of column densities.  The reason is that the resonance line becomes optically thick much faster than the intercombination line.  The ratio $I_{\mathrm{r}}/I_{\mathrm{i}}$ is $\approx23$ times the semi-analytical value at low column densities.  This is because continuum pumping is dramatically enhancing the resonance line relative to the semi-analytical value.  For small column densities, photoexcitation is important and the plasma is in Case-C conditions (see Ferland 1999 and Baker et al. 1938).  For large column densities, the plasma exhibits Case-B behavior (Baker \& Menzel 1938).
The temperature-sensitive ratio $G=(I_{\mathrm{f}}+I_{\mathrm{i}})/I_{\mathrm{r}}$ varies by more than two orders of magnitude when the column density increases from
$N_{\mathrm{H}}=10^{17}$~cm$^{-2}$ to $N_{\mathrm{H}}=10^{24}$~cm$^{-2}$. This trend suggests that the column density must be constrained before using $G$ as an ionization indicator.  In the low column density limit, the $G$-ratio is smaller than the semi-analytical value by a factor of $\approx25$, a finding consistent with the resonance line being enhanced by continuum pumping, as discussed above.

\subsection{Line Ratios as a Function of Ionizing Flux and Volume Density}
Next, we calculate a grid of simulations using the same continuum shape and composition but varying both the flux of hydrogen-ionizing photons and the hydrogen volume density.  We consider two hydrogen column densities, $N_{\mathrm{H}}=10^{18}$~cm$^{-2}$ and $10^{23}$~cm$^{-2}$ to show this dependence.  In Figures \ref{fig:Rcontour}-\ref{fig:Gcontour} we plot a number of O~VII intensity ratios as a function of hydrogen density and ionizing flux.  Note that in a triangular region in the bottom right corner of each contour plot, to the right of $\log n_{\mathrm{H}}=6$ and below $\log \Phi(\mathrm{H})=24$, oxygen is not ionized enough to produce significant helium-like emission.  

\begin{figure}
\centering
\includegraphics[width=6.0in,keepaspectratio=true]{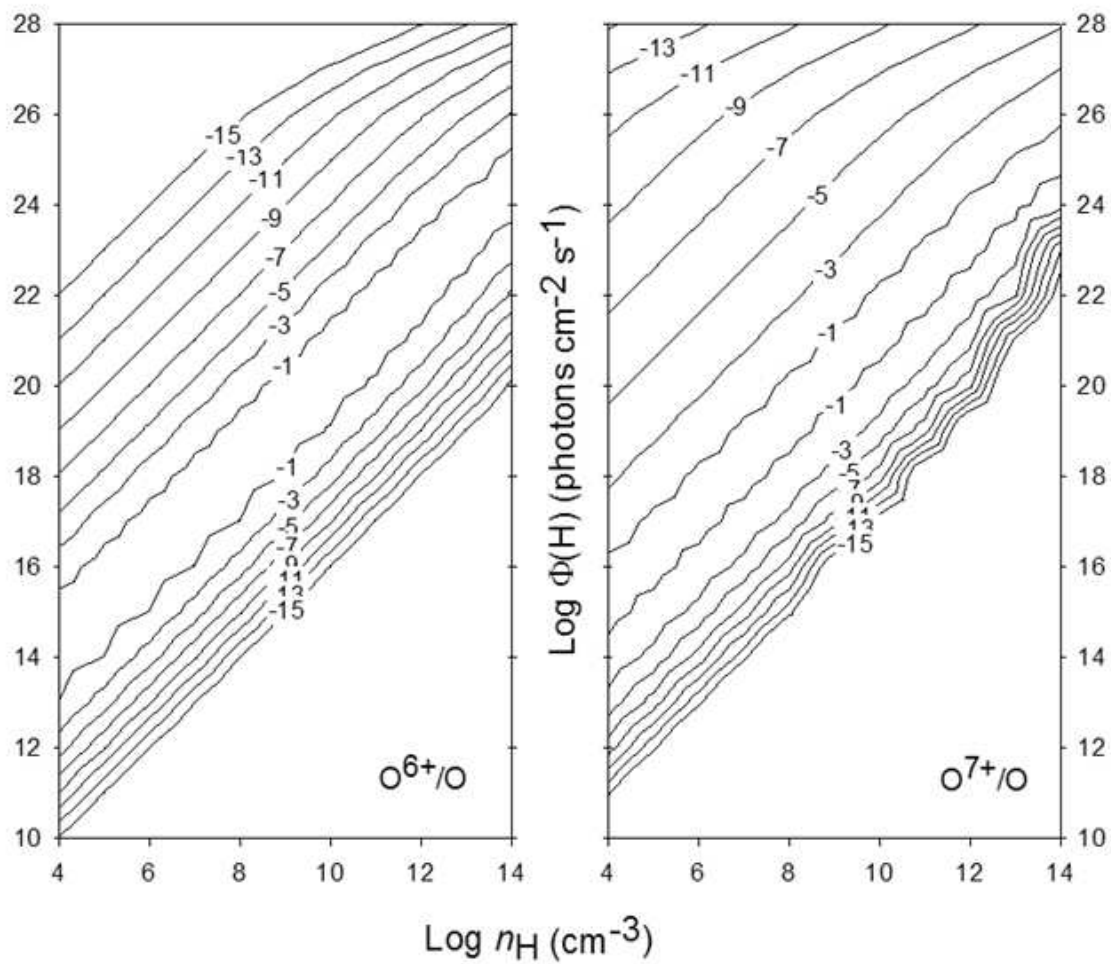}
\caption[Fraction of oxygen in the O$^{6+}$ and O$^{7+}$ ionization states.]{Ionization fraction of oxygen in the (left panel) O$^{6+}$ and (right panel) O$^{7+}$ stages.  Contours are labeled as the log of the fraction.  The column density is $N_{\mathrm{H}}=10^{18}$~cm$^{-2}$ in both panels.  The case with $N_{\mathrm{H}}=10^{23}$~cm$^{-2}$ is very similar except that the lower $10^{-15}$ contour is shifted up nearly a decade for O$^{6+}$ and about half a decade for O$^{7+}$.
Wavy features in this and following contour plots are artifacts in the plotting program due to a sparse grid. }
\label{fig:o6and7plus}
\end{figure}

\begin{figure}
\centering
\includegraphics[width=6.0in,keepaspectratio=true]{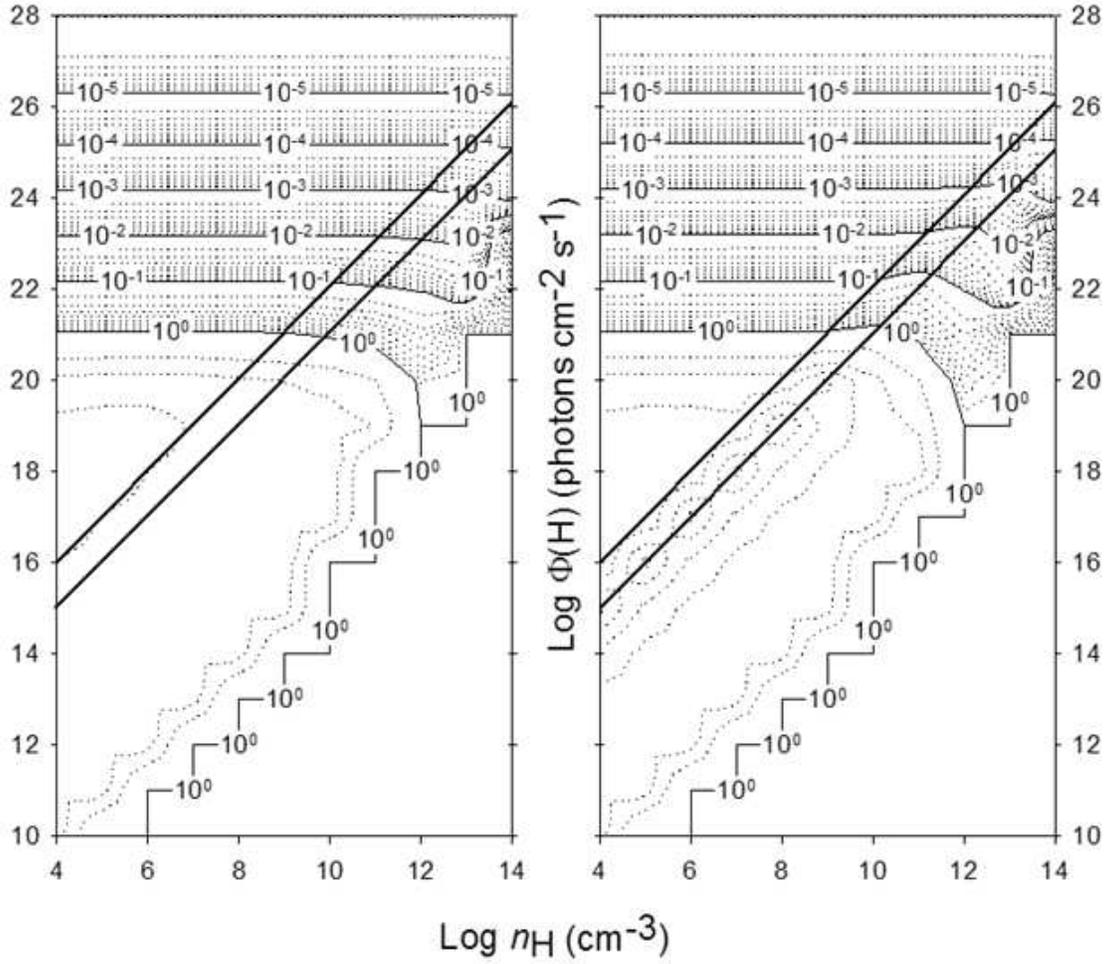}
\caption[Predicted $R=I_{\mathrm{f}}/I_{\mathrm{i}}$ as Function of Hydrogen Density and Ionizing Flux.]{Ratio $R=I_{\mathrm{f}}/I_{\mathrm{i}}$ as a function of hydrogen density and ionizing flux.  Left panel is for a small column density ($N_{\mathrm{H}}=10^{18}$~cm$^{-2}$); right panel is for a large column density ($N_{\mathrm{H}}=10^{23}$~cm$^{-2}$).  The bold diagonal lines in each panel show constant ionization parameter.  See text for details of model.}
\label{fig:Rcontour}
\end{figure}

\begin{figure}
\centering
\includegraphics[width=6.0in,keepaspectratio=true]{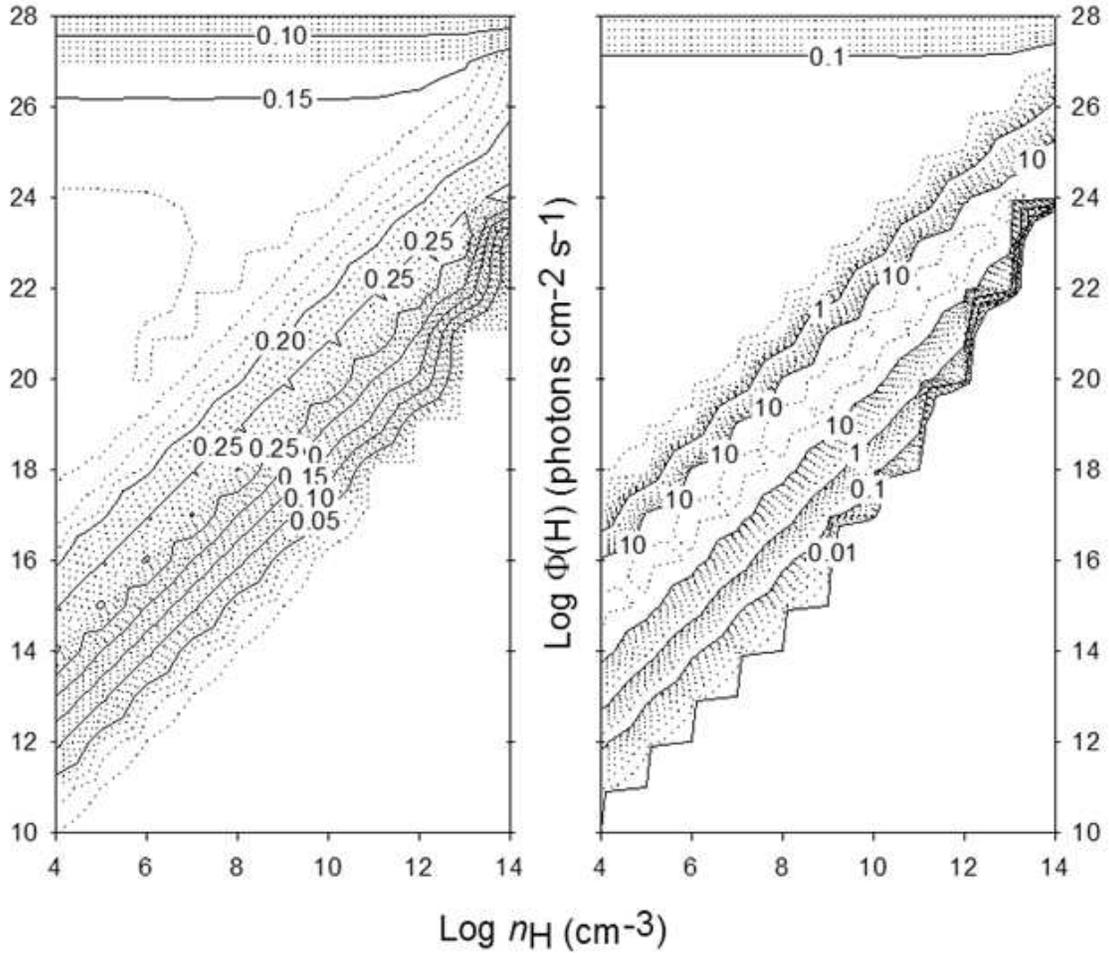}
\caption[Same as Figure~\ref{fig:Rcontour} except ratio is $G=(I_{\mathrm{f}}+I_{\mathrm{i}})/I_{\mathrm{r}}$.]{Same as Figure~\ref{fig:Rcontour} except ratio is $G=(I_{\mathrm{f}}+I_{\mathrm{i}})/I_{\mathrm{r}}$.  Note that the contours have a linear scale in the left panel and a logarithmic scale in the right panel.  When the ionization fractions (Figure~\ref{fig:o6and7plus}) peak, the $G$-ratio is sensitive to ionization parameter and column density.}
\label{fig:Gcontour}
\end{figure}

In Figure~\ref{fig:o6and7plus}, we plot the ionization fractions of O$^{6+}$ and O$^{7+}$.  The plots indicate that in the upper left corner of the plotted parameter space both stages are neglible (because oxygen is almost entirely stripped of electrons).  Moving down and toward the right, the hydrogenic ionization state begins to dominate, followed by the helium-like state.  In the bottom right corner, as noted above, oxygen is too neutral to produce significant helium-like emission.

In Figure~\ref{fig:Rcontour}, we plot the ratio $R=I_{\mathrm{f}}/I_{\mathrm{i}}$ for both column densities mentioned above.  Note that, in both panels, for a given ratio and ionizing flux, there is not a unique density.  Values of $R$ span several orders of magnitude for the entire density range plotted.  This figure, combined with Figure~\ref{fig:randg}, demonstrates that the $R$-ratio is not a simple density diagnostic. Two constant ionization parameter lines are overplotted in each panel in Figure~\ref{fig:Rcontour}.  One can clearly see that, for a given value of $R$, a change in ionization parameter leads to an almost identical change in the derived density.  Considering the $R$-ratio as a function of ionization parameter causes the appearance of a density dependence in the ratio, but that dependence is almost entirely due to the definition of the ionization parameter and not to the physics involved in the $R$-ratio.   The $R$-ratio is roughly constant at low flux, inversely proportional to the flux for several orders of magnitude of the flux, and then constant at very high fluxes.  In the large column density case (right panel), the ratio is roughly a factor of two greater than the corresponding values in the small column density case in the lower left corner of the parameter space.  Values of the O~VII $R$-ratio larger than 4 have been observed but were previously unexplained by theory (see section~\ref{opticaldepth}).
    
The ratio $G=(I_{\mathrm{f}}+I_{\mathrm{i}})/I_{\mathrm{r}}$, a temperature indicator in a collisional plasma, is plotted in Figure~\ref{fig:Gcontour}.  In a photoionized plasma, the ratio is a function of ionization fraction (shown in Figure~\ref{fig:o6and7plus}) and column density.  The ratio depends on both ionization parameter and column density.  This fact can be seen in the right panel of Figure~\ref{fig:Gcontour}, where the column density is large and the $G$-ratio differs greatly from the small column density case in the left panel.

\subsection{Effects of Continuum Pumping}

The results presented above are strongly dependent upon continuum pumping.   This is why the gas reaches Case-C conditions at small column densities (see Figure~\ref{fig:randg}).  Here we further illustrate the effects of continuum pumping by considering a second continuum component.

Some of the above results at high flux remain counterintuitive, in particular that the $R$-ratio in Figure~\ref{fig:Rcontour} should trace ionizing flux (i.e., is parallel to the density axis).  This is caused by direct continuum pumping of the UV3$_j$ lines.  A simple calculation illustrates the effect.  

\begin{figure}
\centering
\includegraphics[width=6.0in,keepaspectratio=true]{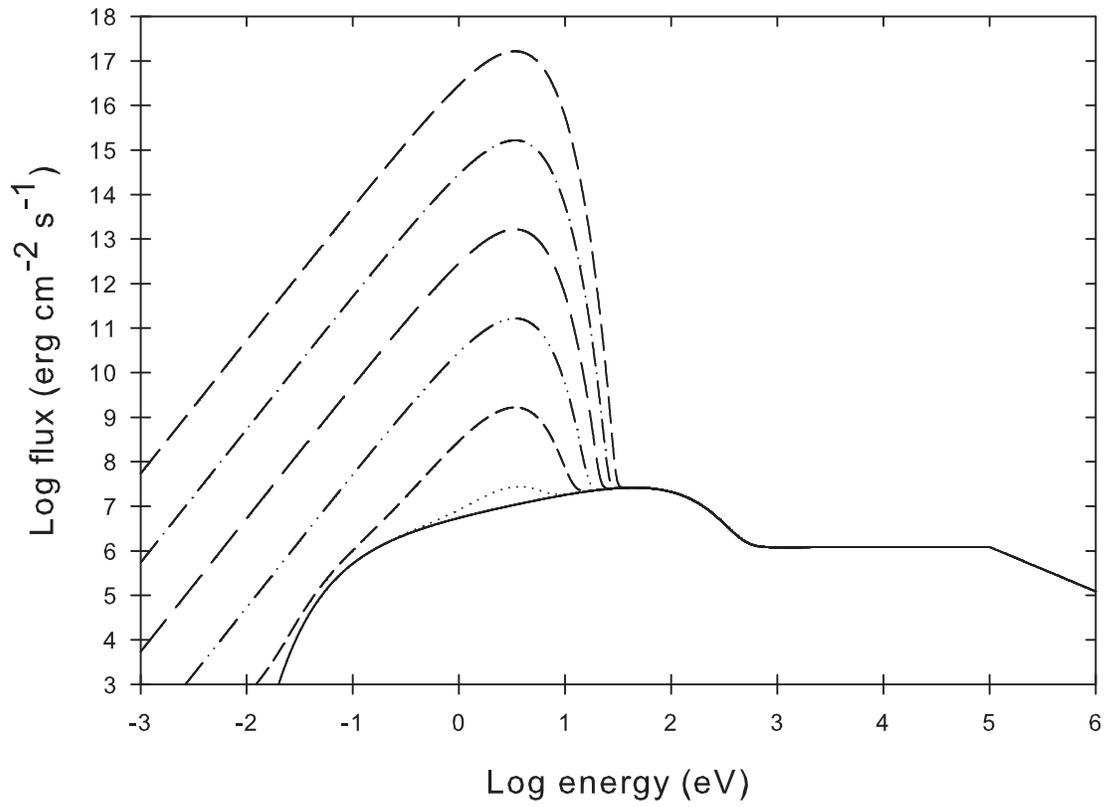}
\caption[The total and individual components of the incident continuum.]{A typical AGN continuum is shown as the solid line.  The soft blackbody, which we vary to show its effect on the X-ray lines, is also shown for several different values of flux.}
\label{fig:incident}
\end{figure}

\begin{figure}
\centering
\includegraphics[width=6.0in,keepaspectratio=true]{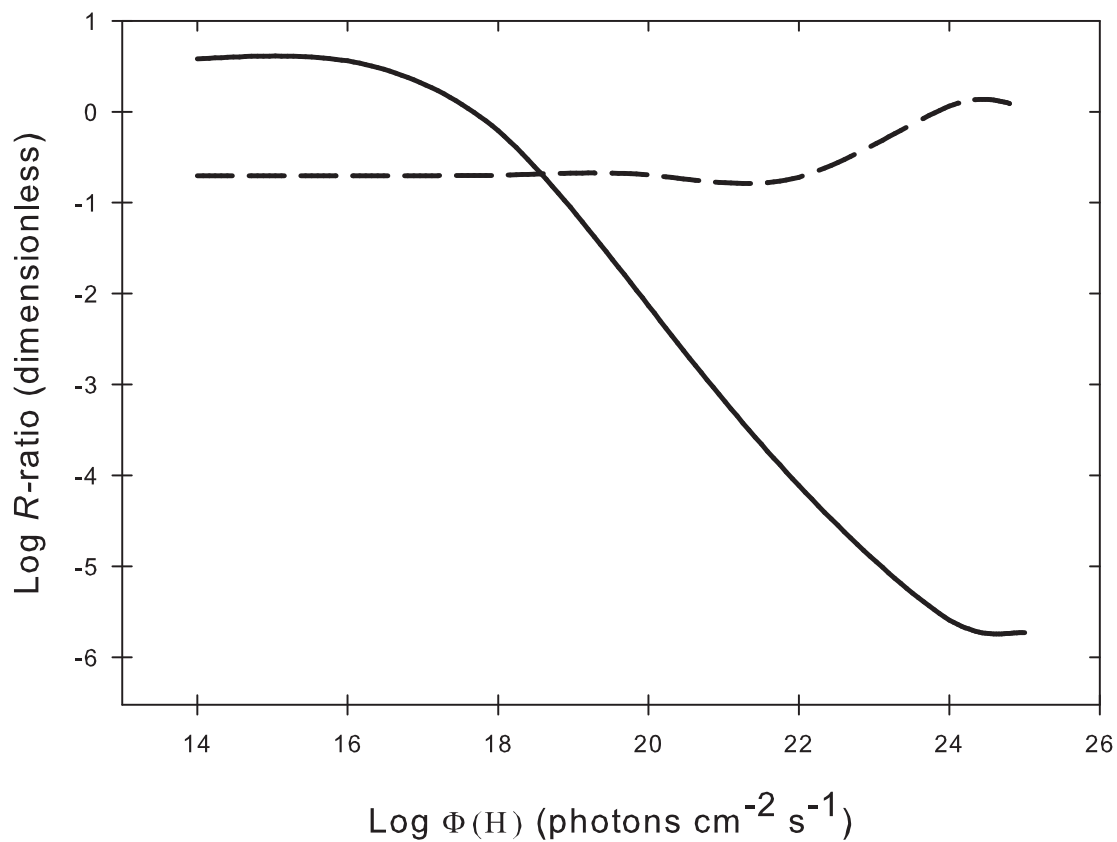}
\caption[$R$-ratio and $G$-ratio as a function of the ionizing flux of an additional continuum component.]{$R$-ratio and $G$-ratio as functions of the flux of the soft component shown in Figure~\ref{fig:incident}.  The low flux limit is the canonical value.  The $R$-ratio high flux limit is where the relative populations of the $2s\,{}^{3}\!S$ and $2p\,{}^{3}\!P_1$ levels are statistically weighted.  The corresponding high flux limit in the $G$-ratio has not been reached because the UV1 line strength is several orders of magnitude less than the resonance line strength.}
\label{fig:Rlimit}
\end{figure}

We add a $10,000$~K blackbody to the original Korista et al. (1997) continuum used above (again with hydrogen-ionizing flux $\Phi(\mathrm{H})=10^{18}$~photons~cm$^{-2}$~s$^{-1}$, hydrogen density $n_{\mathrm{H}}=10^7$~cm$^{-3}$, and hydrogen column density $N_{\mathrm{H}}=10^{18}$~cm$^{-2}$).  The flux of the blackbody is varied.  In Figure~\ref{fig:incident}, we plot the Korista continuum and the soft blackbody with several values of flux.  We understand that the net continuum shown in Figure~\ref{fig:incident} is not representative of any real continuum; this exercise is simply meant to demonstrate the effect of the continuum at $\approx1640~\mbox{\AA}$, the wavelength of the UV3$_j$ transitions (Table~\ref{table:wavelengths}), upon the X-ray line ratios.  In Figure~\ref{fig:Rlimit}, we plot the $R$-ratio as a function of the flux of the second continuum component, holding everything else constant.  The ratio changes by several orders of magnitude.  The blackbody only contributes to the net continuum for energies less than about 10-30~eV.  This is much less than the 165~eV needed to ionize ions in the $n=2$ shell of O$^{6+}$ and the 560~eV needed to ionize ions in the ground state.  The blackbody continuum is, however, hard enough to directly pump the UV3$_j$ lines.  This pumping changes the relative populations of the $2s\,{}^{3}\!S$ and $2p\,{}^{3}\!P_j$ levels, which in turn changes the $R$-ratio.

For large values of the flux of the blackbody, the ratio goes to a well-defined limit as the relative populations of $2s\,{}^{3}\!S$ and $2p\,{}^{3}\!P_j$ levels become populated according to statistical weights.  In this limit, $n_{2s\,{}^{3}\!S}\approx n_{2p\,{}^{3}\!P_1}$, and 
since $\lambda_{f} \approx \lambda_{i}$, Equation~\ref{eqn:Eratiof} reduces to 
\begin{equation}
R = \frac{A_{2s\,{}^{3}\!S-1s\,{}^{1}\!S}}{A_{2p\,{}^{3}\!P_1-1s\,{}^{1}\!S}} = \frac{1.0\times10^{3}~s^{-1}}{5.5\times10^8~s^{-1}} = 1.8\times10^{-6}.
\end{equation}
This limit is also being approached in the extreme upper portion of Figure~\ref{fig:Rcontour}, in calculations with very high ionizing flux  for similar reasons.

We also plot the $G$-ratio in Figure~\ref{fig:Rlimit}.  Pumping in the singlet levels by the continuum at UV1 is much less effective than in the triplet case because the UV1 line strength is many orders of magnitude less than the resonance line strength, especially considering the fact that the resonance line is already being pumped in the low flux limit.  Because the resonance intensity (the $G$-ratio denominator) is not strongly affected by the UV1 pumping, and because the UV3$_j$ pumping decreases one term of the $G$-ratio numerator while increasing the other, the $G$-ratio is much less affected than the $R$-ratio.  The $G$-ratio will finally go to a high-flux limit at fluxes roughly three orders of magniude higher than those plotted in Figure~\ref{fig:Rlimit}.

\section{Conclusions}
\label{conclusions}

\begin{description}
\item[] The $R$-ratio measures the flux in the continuum in a photoionized plasma.  It is a good density indicator only for a known ionization fraction.  The ratio is strongly dependent upon continuum pumping of the UV3$_j$ lines between $n=2$ levels.  Because of this dependence, X-ray observations can be used to deduce the UV radiation field striking the gas where the lines form.  This is important for AGN since several components of the continuum are beamed, and the continuum we observe may not reflect the continuum striking clouds.  

\item[] Optical depth in the intercombination line causes the high observed values of the $R$-ratio.  This was previously unexplained.  Caution must be taken in using $R$ as a density indicator in conditions where the intercombination line may be optically thick.

\item[] While the $G$-ratio does track ionization fraction well, the column density must be constrained before using $G$ as a ionization indication, and should not be used as a temperature indicator at all in a photoionized plasma.

\item[] The resonance line is directly pumped by the incident continuum for small column densities.  This affects the $L$ and $G$ ratios by large amounts.

\item[] Absorption of UV3$_j$ lines may occur in some environments with large column densities.

\item[] For some helium-like species, the optical depth of UV3$_j$ lines (or column density of the ion) may be determined from X-ray observations of that ion's intercombination lines, provided the intercombination doublet can be resolved.  The column density could be deduced from this. 

\end{description}

\acknowledgements
We thank Jack Baldwin and Andy Fabian for helpful discussions, and the referee, Manuel Bautista, for an insightful review.  We also thank NSF for support through AST-0607028 and NASA for support through NNG05GG04G.

\appendix
\section{Data for Helium-like Ions}
\label{AppendixData}

Here we discuss the sources and methods of our theoretical models.  Where details are omitted, the model ion is treated in the same way as the model helium atom (Porter et al. 2005, Bauman et al. 2005).

\subsection{Energies}
For level energies up to and including $n=5$, we use the energies from version 5.0 of Chianti (Dere et al. 1997, Landi et al. 2006).  Where Chianti does not provide an energy we calculate energies using fits to quantum defects and Equation~1 of Bauman et al. (2005), modified to include the multiplicative factor $(Z-1)^2$ in the numerator.  This multiplicative factor represents the approximate scaling with $Z$ of ionization energy of an ion in the ground state.  The one is subtracted from the nuclear charge because the charge ``seen'' by an excited electron is partially screened by the charge of the second electron (in the ground state).  In this formalism, the quantum defect is nearly independent of $Z$.

\subsection{Collisional Data}
For electron impact collisions from ground to $n=2$ levels and between the $n=2$ levels, we use simple fits to the effective collision strengths of Zhang \& Sampson (1987).  The most accurate data for these transitions (for oxygen) are from Delahaye \& Pradhan (2002) who say their results generally agree well with the Zhang \& Sampson results.  For Stark ($l$-mixing) collisions, we use the method of Seaton (1962) for $l\leq2$ and Vrinceanu \& Flannery (2001) for greater $l$.  We apply the method of virtual quanta (Jackson 1999) to calculate $l$-mixing collision rates due to proton and singly-ionized helium impact.  (These collisions are treated in the same way as in the case of helium, described in Porter et al. 2005, where they are discussed in some detail.)  For $n$-changing collisions, we use the method of Vriens \& Smeets (1980), Equations~14-16.  We apply the method of virtual quanta to the Vriens \& Smeets method in order to calculate $n$-changing collision rates due to proton and singly-ionized helium impact.  (It is worth noting, however, that electron impact excitation usually completely dominates over proton impact for $\Delta n > 0$ transitions.)  Collisional de-excitation is included via detailed balancing.  For collisional ionization, we take the minimum positive result from the hydrogenic methods of Allen (1973) and Sampson \& Zhang (1988).  Proton impact excitations are negligible for oxygen but increasingly important for increasing nuclear charge (i.e., our rates generally agree with the trend reported by Blaha [1971]).  As a basic test of our results in a collisional plasma, we note that our O~VII $R$-ratio curve at 500,000~K has very nearly the same density dependence as is seen in Figure~3 of Bautista \& Kallman (2000) and Figure~8 of Porquet \& Dubau (2000).  

\subsection{Recombination Coefficients}
We add the results of equation 15 of Seaton (1959) for $101\leq n\leq 1000$ to the direct recombination into the highest collapsed level at $n=100$.  The recombination coefficients into the terms with $n\leq100$ are calculated using the Milne relation.  For $n\leq10$ we use fits to TOPbase (Fernley et al. 1987; Cunto et al. 1993) photoionization cross-sections for the least hydrogenic $S$ terms.  We use hydrogenic cross-sections for the rest.  Photoionization from all levels is included, as is induced recombination via detailed balancing.

State-specific dielectronic recombination coefficients for $n\le 8$ and $l\le 4$ are interpolated from Badnell (2006).  Rates to levels with $n>8$ are assumed to follow a $n^{-3}$ law, while rates to levels with $l>4$ are neglected.  We make the simple approximation that all doubly-excited levels have the same energy, equal to the energy of the $n=2$ shell of the H-like stage, so that satellite line intensities are proportional to the state-specific dielectronic recombination coefficients.  We intend to improve this treatment in future work, but the current approximation is sufficient for photoionized plasma calculations.  We note that Bautista \& Kallman (2000) warn that satellite lines lie close in wavelength to the forbidden and intercombination lines and will enhance their apparent intensities in low and medium resolution spectra.

\subsection{Transition Probabilities}

\subsubsection{Allowed Transitions}
We use the transition probabilities of Johnson et al. (2002) where available.  For $n=2$ to $n=2$ transitions not available in Johnson et al, we use simple fits to data from the NIST Atomic Spectra Database, version 3.0.3\footnote[3]{see http://physics.nist.gov/PhysRefData/ASD/index.html} for $2p\,{}^{3}\!P_j$-$2s\,{}^{3}\!S$ and TOPbase for $2p\,{}^{1}\!P$-$2s\,{}^{1}\!S$.  We also use TOPbase for same-$n$ transitions with $n>2$.  Then, for transitions with either initial or final $l>2$, we use the hydrogenic formula given by Drake (1996).  Next, transitions to ground, $2s\,{}^{1}\!S$, $2s\,{}^{3}\!S$, $3s\,{}^{3}\!S$, or $4s\,{}^{3}\!S$ are calculated via extrapolation of Johnson et al. values to higher initial $n$.  All other allowed transitions are calculated through the use of Drake's semi-classical quantum-defect routine.

\subsubsection{Forbidden Transitions}
The transition probabilities for the transitions $2s\,{}^{3}\!S-1s\,{}^{1}\!S$ are calculated from a fit to Lin et al. (1977) values.  Two-photon transition probabilities are taken from Derevianko \& Johnson (1997).  The distribution of photon energies are from Johnson (2002).  The sources of intercombination transition probabilities are as follows:
upper level $2p\,{}^{3}\!P_1$: fits to Johnson et al. (2002) for $Z\leq18$, and Lin et al. (1977) otherwise; and
upper level $2p\,{}^{3}\!P_2$: fits to Lin et al. (1977).  Transition probabilites for transitions $np\,{}^{3}\!P-n's\,{}^{1}\!S$ and $np\,{}^{1}\!P-n's\,{}^{3}\!S$ with $n\neq n'$ are taken from fits to (or extrapolation of) Johnson et al. values.  Finally, transition probabilities for the transition $2p\,{}^{1}\!P-2s\,{}^{3}\!S$ are taken from fits to values in Savukov et al. (2003).

\section{Absorption of Intercombination and UV3$_j$ Lines}
\label{AppendixAbsorption}
It is not generally known that absorption of the intercombination and UV3$_j$ lines of some helium-like ions may be possible in some environments.  Here we present arguments in this regard and derive column densities necessary to detect the absorption of these lines.

The line-absorption coefficient of a line with lower and upper levels $l$ and $u$ and central frequency $\nu_0$, neglecting radiative damping, is
\begin{equation}
\kappa_\nu = \frac{\sqrt{\pi}e^2}{m_ec}\frac{f_{lu}}{\Delta\nu_D}\exp\left[-\left(\frac{\nu-\nu_0}{\Delta\nu_D}\right)^2\right]\ \ [\mathrm{cm}^{2}],
\label{eqn:kappa}
\end{equation}
where $e$ is the electronic charge, $m_e$ is the electronic mass, $c$ is the speed of light, $f_{lu}$ is the absorption oscillator strength of the line, and $\Delta\nu_D$ is the Doppler width [Hz] (which consists of thermal and turbulent components added in quadrature).  Equation \ref{eqn:kappa} reduces, at line center, to
\begin{equation}
\kappa_0 = 0.0150~\frac{f_{lu}(Z)}{\Delta\nu_D}\ \ [\mathrm{cm}^{2}].
\label{eqn:kappa2}
\end{equation}
The optical depth at line center is
\begin{equation}
\tau_0 = \int \kappa_0~n_{l}(Z,T)~dl\ \ [\mathrm{dimensionless}],
\label{eqn:tau}
\end{equation}
where $n_{l}(Z,T)$ is the population of level $l$ and the integral is over the length of the absorbing region.  

For the $2p\,{}^{3}\!P_j-2s\,{}^{3}\!S$ UV3$_j$ transitions of helium-like ions (with wavelengths given in Table~\ref{table:wavelengths}), 
we substitute Equation~\ref{eqn:2s3Spop} into equation \ref{eqn:tau} and, assuming an isothermal absorbing region, obtain 
\begin{equation}
\tau_0 \approx \kappa_0~\frac{\alpha^{eff}_{2s\,{}^{3}\!S}(Z,T)}{A_{2s\,{}^{3}\!S-1s\,{}^{1}\!S}(Z)}~\int n_e~n_{\mathrm{Z}^+}~dl.
\label{eqn:tau2}
\end{equation}
If we further assume that the electron density does not change too much over the length of the absorbing region, then
\begin{equation}
\tau_0 \approx \kappa_0~\frac{\alpha^{eff}_{2s\,{}^{3}\!S}(Z,T)}{A_{2s\,{}^{3}\!S-1s\,{}^{1}\!S}(Z)}~n_e~N_{\mathrm{Z}^+},
\label{eqn:tau3}
\end{equation}
where $N_{\mathrm{Z}^+}$ is the column density (cm$^{-2}$) of the hydrogen-like ion.

For oxygen at $500,000$~K, we find $\alpha^{eff}_{2s\,{}^{3}\!S}(8, 10^{5.7}$~K$) \approx 1.1\times10^{-12}$~cm$^{3}$~s$^{-1}$ and, neglecting turbulence, $\Delta\nu_D = 1.4\times10^{11}$Hz, so that $\tau_0 \approx 5.7\times10^{-30}~n_e~N_{\mathrm{O}^{7+}}$ for the strongest line of the triplet.  For appreciable absorption, letting $\tau_0 = 0.1$, we need $n_e~N_{\mathrm{O}^{7+}} \geq 2\times10^{28}$~cm$^{-5}$.  At $n_e=10^{10}$~cm$^{-3}$ one would need $N_{O^{7+}}\approx 2\times10^{18}$~cm$^{-2}$ to detect absorption.  That column density is one to two orders of magnitude larger than is typically reported but may exist in some environments.  A larger volume density would allow for a smaller column density.  Absorption is thus likely to be detectable in some environments.  The likelihood of detection decreases, however, with increasing nuclear charge for two reasons.  First, the transition probability in the denominator of Equation~\ref{eqn:tau3} increases much faster than any other factors change.  Second, astrophysical abundances tend to decrease dramatically for elements heavier than oxygen.  Absorption of the UV3$_j$ lines will be considered in a future paper.

For the $2p\,{}^{3}\!P_j-1s\,{}^{1}\!S$ transitions of helium-like ions, we can make the approximation that all ions are in the ground state and obtain
\begin{equation}
\tau_0 \approx \kappa_0~N_{\mathrm{Z}},
\end{equation}
where $N_{\mathrm{Z}}$ is the column density (cm$^{-2}$) of the helium-like ion.
Again considering oxygen at $500,000$~K, we find $\Delta\nu_D = 1.1\times10^{13}$Hz, so that $\tau_0 \approx 2.1\times10^{-19}~N_{\mathrm{O}^{6+}}$ for the strongest line of the doublet, where we have used $f_{1s\,{}^{1}\!S,2p\,^{3}\!P_1}=1.2\times10^{-4}$.  Again letting $\tau_0 = 0.1$, we need $N_{\mathrm{O}^{6+}} \geq 5\times10^{17}$~cm$^{-2}$.  If only one-tenth of the oxygen in the column is helium-like oxygen, and the oxygen abundance was comparable to solar, the total hydrogen column density necessary to see significant absorption of the O~VII intercombination line would be about $10^{22}$~cm$^{-3}$.  Absorption is thus likely to be detectable in some environments, which has been stated by Godet et al. (2004), and may have been detected (see Kinkhabwala et al. 2002).  Including turbulence will reduce the optical depth and increase the necessary optical depth to observe absorption (although the effects will be strongly dependent upon line profile).  The likelihood of detection decreases with increasing nuclear charge, however, because the intercombination rate increasingly dominates the UV3$_j$ rates (so that the intercombination lines effectively become the only decay mode from $2p\,{}^{3}\!P_j$), and because abundances tend to decrease dramatically for elements heavier than oxygen.  

\clearpage

\begin{deluxetable}{lrccccccc}
\rotate
\tabletypesize{\scriptsize}
\tablecaption{Wavelengths of the UV, Intercombination, Forbidden, and Resonance Transitions of He-like Ions.}{}
\tablewidth{0pt}
\tablehead
{
  \colhead{}     & 
  \colhead{}                             & 
  \multicolumn{7}{c}{Wavelength ($\mbox{\AA}$)}		\\
  \colhead{Chemical}     & 
  \colhead{}                             & 
  \colhead{UV3$_0$}						&
  \colhead{UV3$_1$}						&
  \colhead{UV3$_2$}						&
  \colhead{UV1}						  &
  \colhead{i}						  &
  \colhead{f}             &
  \colhead{r}							\\					
  \colhead{Symbol}     & 
  \colhead{Z}                             & 
  \colhead{$2p\,{}^{3}\!P_0-2s\,{}^{3}\!S$}						&
  \colhead{$2p\,{}^{3}\!P_1-2s\,{}^{3}\!S$}						&
  \colhead{$2p\,{}^{3}\!P_2-2s\,{}^{3}\!S$}						&
  \colhead{$2p\,{}^{1}\!P-2s\,{}^{1}\!S$}						  &
  \colhead{$2p\,{}^{3}\!P-1s\,{}^{1}\!S$}						  &
  \colhead{$2s\,{}^{3}\!S-1s\,{}^{1}\!S$}             &
  \colhead{$2p\,{}^{1}\!P-1s\,{}^{1}\!S$}												
}
\startdata
C	&	6	&	2277.2	$\;\:\:$	&	2277.8	$\;\:\:$	&	2270.8	$\;\:\:$	&	3526.7	$\;\:\:$	&	40.731	$\;\:\:$	&	41.472	$\;\:\:$	&	40.268	$\;\:\:$	\\
N	&	7	&	1907.6	$\;\:\:$	&	1907.3	$\;\:\:$	&	1896.7	$\;\:\:$	&	2896.4	$\;\:\:$	&	29.084	$\;\:\:$	&	29.534	$\;\:\:$	&	28.787	$\;\:\:$	\\
O	&	8	&	1639.9	$\;\:\:$	&	1638.3	$\;\:\:$	&	1623.6	$\;\:\:$	&	2449.0	$\;\:\:$	&	21.807	$\;\:\:$	&	22.101	$\;\:\:$	&	21.602	$\;\:\:$	\\
F	&	9	&	1417.2	$\;\:\:$	&	1414.2	$\;\:\:$	&	1395.5	$\;\:\:$	&	2136.1	$\;\:\:$	&	16.947	$\;\:\:$	&	17.153	$\;\:\:$	&	16.807	$\;\:\:$	\\
Ne	&	10	&	1277.7	$\;\:\:$	&	1272.8	$\;\:\:$	&	1248.3	$\;\:\:$	&	1856.0	$\;\:\:$	&	13.553	$\;\:\:$	&	13.699	$\;\:\:$	&	13.447	$\;\:\:$	\\
Na	&	11	&	1149.2	$\;\:\:$	&	1142.3	$\;\:\:$	&	1111.8	$\;\:\:$	&	1646.9	$\;\:\:$	&	11.083	$\;\:\:$	&	11.192	$\;\:\:$	&	11.003	$\;\:\:$	\\
Mg	&	12	&	1043.3	$\;\:\:$	&	1034.3	$\;\:\:$	&	997.46		&	1474.2	$\;\:\:$	&	9.2312		&	9.3143		&	9.1688		\\
Al	&	13	&	954.33		&	943.16		&	899.67		&	1327.7	$\;\:\:$	&	7.8070		&	7.8721		&	7.7573		\\
Si	&	14	&	878.65		&	865.14		&	814.69		&	1200.7	$\;\:\:$	&	6.6883		&	6.7404		&	6.6480		\\
S	&	16	&	756.31		&	738.32		&	673.40		&	991.95		&	5.0665		&	5.1015		&	5.0387		\\
Ar	&	18	&	661.56		&	639.55		&	559.97		&	823.25		&	3.9691		&	3.9939		&	3.9488		\\
Ca	&	20	&	585.93		&	560.74		&	466.90		&	687.95		&	3.1928		&	3.2111		&	3.1772		\\
Fe	&	26	&	428.23		&	400.30		&	271.16		&	382.76		&	1.8595		&	1.8682		&	1.8504		\\
\enddata
\label{table:wavelengths}
\end{deluxetable}

\clearpage

\clearpage

\clearpage

\clearpage

\clearpage

\clearpage

\clearpage

\end{document}